# Cavity enhanced emission from a silicon T center

*Fariba Islam,*[†,‡,#] *Chang-Min Lee,*[†,‡,#] *Samuel Harper,*[†,‡,] *Mohammad Habibur Rahaman,*[†,‡,] *Yuqi Zhao,*[†,‡,] *Neelesh Kumar Vij,*[†,‡,] *and Edo Waks*[†,‡,*]

[†]Institute for Research in Electronics and Applied Physics and Joint Quantum Institute, University of Maryland, College Park, Maryland 20742, USA

[‡]Department of Electrical and Computer Engineering, University of Maryland, College Park, Maryland 20740, USA

**Abstract:**

Silicon T centers present the promising possibility to generate optically active spin qubits in an all-silicon device. However, these color centers exhibit long excited state lifetimes and a low Debye-Waller factor, making them dim emitters with low efficiency into the zero-phonon line. Nanophotonic cavities can solve this problem by enhancing radiative emission into the zero-phonon line through the Purcell effect. In this work we demonstrate cavity-enhanced emission from a single T center in a nanophotonic cavity. We achieve a two-orders of magnitude increase in brightness of the zero-phonon line relative to waveguide-coupled emitters, a 23% collection efficiency from emitter to fiber, and an overall emission efficiency into the zero-phonon line of 63.4%. We also observe a lifetime enhancement of 5, corresponding to a Purcell factor exceeding 18 when correcting for the emission to the phonon sideband. These results pave the way towards efficient spin-photon interfaces in silicon photonics.

Silicon photonic devices integrate many optical components into a compact chip through foundry-compatible fabrication processes, making them an attractive choice for quantum information applications[1]. Color centers in silicon have recently emerged as promising quantum light sources that can be directly engineered in this highly scalable photonic platform. Significant progress has been made in the development of quantum light sources based on various silicon centers including the G center[2–7] and W center[7–10]. But these emitters do not possess an unpaired electron, and therefore lack the ability to act as spin-based quantum memories essential for quantum networking[11] and distributed and modular quantum computing[12,13].

Unlike the G and W centers, the T center does possess an unpaired electron, enabling it to function as an optically active spin qubit. It also exhibits remarkably long coherence times that can exceed 1 ms[14,15]. In addition, it emits at the telecom O-band with a narrow emission linewidth[16], making it compatible with existing fiber infrastructure for long-distance quantum communication. Recent work has successfully isolated single T centers,[14] representing a significant milestone in the field. Subsequent work employed waveguides to isolate single T centers with improved efficiency[17,18]. But in order to utilize these promising spin-qubit systems for quantum applications, significant challenges must be overcome. In contrast to the G and W centers, the T center has a long excited state lifetime of approximately 1 μs, which makes it a dim emitter. Furthermore, it exhibits a Debye-Waller factor of only 0.23, which means the majority of photons are lost to the broad and incoherent phonon sideband resulting in a low radiative efficiency into the zero-phonon line.

Nanophotonic cavities offer a potential solution to the low brightness and poor radiative efficiency found in T centers. These cavities can enhance the spontaneous emission rate through the Purcell effect[19], improving radiative efficiency into the zero phonon line. This strategy has been effectively employed in many quantum emitters including quantum dots[20–22], color centers in diamond[23,24] and SiC[25], and rare-earth ions[26–31]. It has also already been employed with G centers[32–34]. But the integration of T centers with nanophotonic cavities for strong radiative enhancement remains to be demonstrated, and represents an important step towards utilizing these spin qubits for practical quantum applications.

In this letter we report strong radiative enhancement of a single silicon T center in a nanophotonic cavity. We employ a nanobeam cavity structure that both enhances spontaneous emission and allows efficient extraction of photons into a single mode fiber. Using this approach, we demonstrate a two orders of magnitude improvement in the brightness of the zero-phonon line of a single T center, a photon collection efficiency from emitter into fiber exceeding 23%, and a total emission efficiency into the zero-phonon line of 63.4%. From time-resolved lifetime measurements, we observe a lifetime enhancement of 5, which corresponds to a Purcell factor exceeding 18 for the zero-phonon line when correcting for incoherent emission to the phonon sideband. These results represent an important step towards developing high-efficiency spin-light interfaces using silicon color centers.

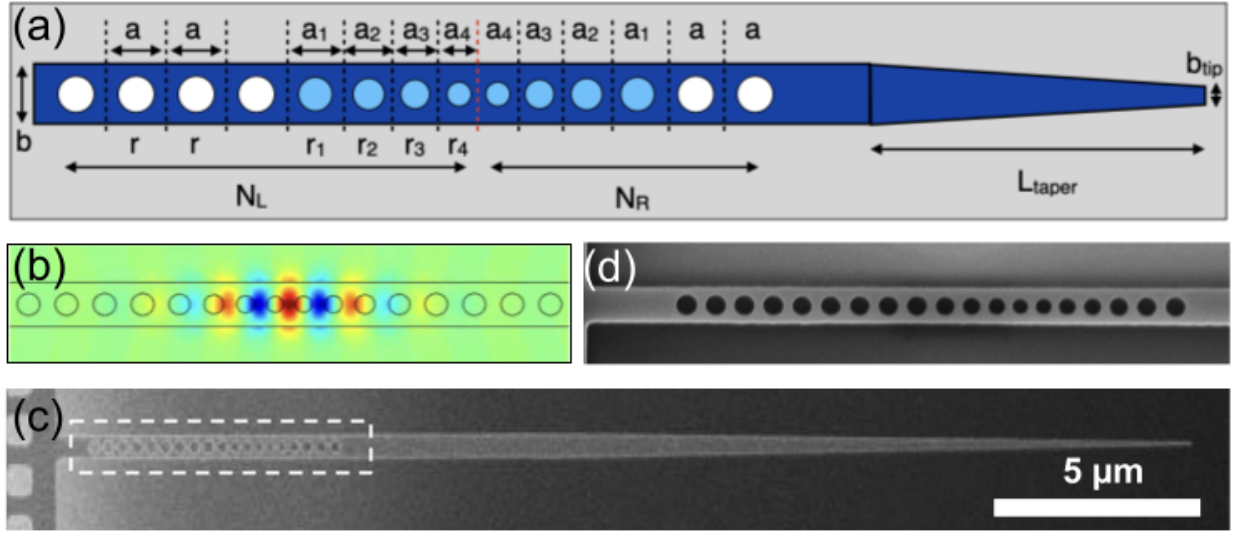

**Figure 1.** (a) Schematic of the nano-cavity design. The photonic crystal cavity is formed from a 4-hole linear-taper defect in a periodic hole array. The parameters of the cavity are the number of holes on the left ($N_L$) and right ($N_r$), the periodicity of the photonic crystal hole array ($a$) and taper ($a_i$), and the hole radii of the mirror ($r$) and taper ($r_i$), and the length of the nanobeam taper ($L_{taper}$). (b) Simulated cavity mode electric field intensity. (c) Scanning electron microscope (SEM) image of the fabricated cavity and nanobeam taper. (d) Close-up SEM image of the white dashed box in panel c, showing the cavity region.

Figure 1(a) illustrates the nanobeam photonic crystal cavity design. This device comprises a one-dimensional air-clad silicon nanobeam photonic crystal cavity with an adiabatic waveguide taper on one side, which allows for outcoupling to a lensed fiber. The photonic crystal cavity is composed of an array of air holes, with the cavity region formed by linearly tapering both the hole radii and periodicity[35]. The number of holes to the left ($N_l$) and right ($N_r$) can be individually selected to control the in-plane coupling in either direction. In our device we employed a 4-hole taper. We optimized the cavity using commercial finite-difference time-domain simulation software to achieve a high $Q$ mode at the T center resonance. From these simulations we determined an optimal value of $a = 375\ nm$, and $r = 120\ nm$, with the hole radii in the cavity region tapered from 86 nm to 112 nm, while the hole periodicities were tapered from 270 nm to 349 nm. Our design used an $N_r = 6$ and $N_l = 13$ to ensure the majority of the light is emitted to the tapered region. Figure 1(b) shows the electric field intensity of the optimized mode. The designed quality factor was $Q = 11700$, dominated by in-plane losses to the mode propagating towards the tapered nanobeam. We note that, from experimental measurements, we observed a 30 nm blueshift of the cavity mode relative to the calculated resonance. For this reason we designed the cavity to be resonant at 1356 nm, resulting in an actual cavity resonant frequency that matches the T center emission.

To create the nanobeam structure, we employed a combination of electron beam lithography and transfer print lithography techniques. The fabrication method and T center implantation approach we used is identical to the one we previously reported for fabrication of bare tapered nanobeams[18]. Figure 1c shows a fabricated nanobeam cavity suspended from the edge of a silicon carrier chip. Figure 1d shows a close-up image of the dashed box in panel c which contains the cavity region of the device.

All measurements were performed in a cryogenic fiber probe station with an integrated magnetic field. We placed the fiber probe station in a closed cycle refrigerator that cooled the sample to 3.6 K. An integrated Helmholtz coil magnet could apply up to 7 T of magnetic field in the plane of the sample. Sample excitation was performed from the out-of-plane direction using a low-temperature microscope objective using a 794 nm laser. We collected the emission from the cavity using a lensed fiber whose numerical aperture was matched to the nanobeam. We performed spectral measurements using a grating spectrometer coupled to an InGaAs CCD array, with a spectrometer resolution of 0.03 nm. Photon counting measurements were performed using superconducting nanowire detectors with quantum efficiency of 0.88. For photon counting measurements we filtered all emission using a tunable fiber Fabry-Perot grating with a spectral width of 0.02 nm, in order to filter out only the zero-phonon line.

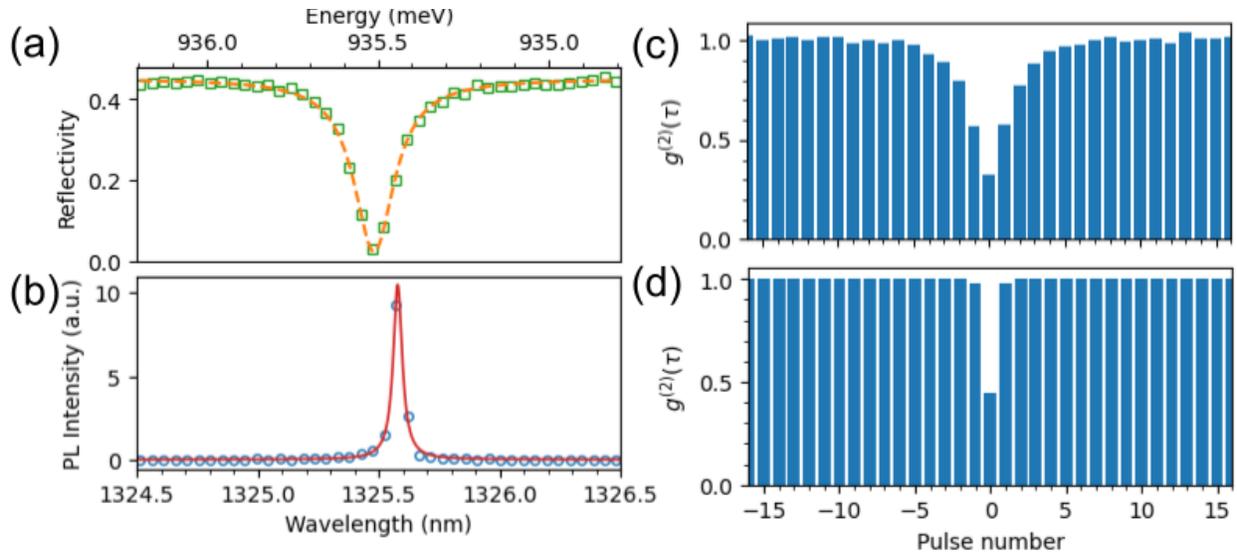

**Figure 2.** (a) Reflectivity measurement of the cavity. Green squares represent the measured data and the dashed line is a numerical fit to a Lorentizian function. From the numerical fit we determine the cavity quality factor to be $Q = 6600$, and the cavity resonant wavelength to be $\lambda_c = 1325.48\ nm$. (b) Photoluminescence measurement of the cavity showing a bright spectrometer resolution limited peak corresponding to the zero-phonon line of a T center. The blue circles are measured data and the solid red line is a numerical fit. From the fit we determine the center wavelength of the zero-phonon line to be $\lambda_z = 1325.58\ nm$. (c)-(d), Second-order autocorrelation histogram of the zero-phonon line using a pulsed excitation laser at an average pump power of 10 nW (c) and 370 nW (d).

We first characterized the bare cavity through direct reflectivity measurements. Figure 2a shows the reflection spectrum, taken using a broad-band Tungsten-Halogen lamp. The reflectivity exhibits a sharp dip at the cavity resonant wavelength of 1325.48 nm. On-resonance, the dip is almost fully suppressed, indicating that the cavity is operating near critical coupling. Under this condition, half of the light emits into the linear taper region, while the remainder is lost to leaky free-space modes and optical absorption. The solid line represents a Lorentzian fit of the cavity spectrum. From the linewidth of the fit we determined a cavity quality factor of $Q = 6600$. By directly measuring the reflected power from the cavity off-resonance, we calculated the coupling efficiency between the nanobeam and the lensed fiber to be 73% (see Supporting Information section 1).

We next performed photoluminescence measurements on the structure. Figure 2b shows the photoluminescence spectrum. We excited the sample using a continuous wavelength laser emitting at a wavelength of 780 nm with average power of 500 nW (measured before the optical window of the cryostat). The photoluminescence spectrum exhibits a sharp peak near the cavity resonance, attributable to a T center. The solid line represents a Lorentzian fit. From the fit, we determined the center wavelength of the emitter to be 1325.58 nm, which is nearly resonant with the cavity.

To validate that the line corresponds to a single T center, we performed second-order correlation measurements. We excited the T center with 80 ns optical pulses at a repetition period of 5 μs generated by sending the pump laser into a fiber-coupled acousto-optic modulator with a response time of 5 ns. We used the fiber Bragg grating to isolate the zero-phonon line emission and filter out all other wavelengths. The filtered emission was sent directly to photon counters.

Figure 2c shows the measured second-order correlation $g^{(2)}(\tau)$ of the zero-phonon line emission, taken at an average pump power of 10 nW. The bar graph plots the total counts integrated over a 5 μs window around each peak of the correlation measurement. The second-order correlation shows a clear suppression at $\tau = 0$, with $g^{(2)}(0) = 0.322$. This value is below 0.5, ensuring that the sharp emission peak corresponds to a single T center emitter. We note that all second-order correlation measurements are raw data without background subtraction.

In addition to anti-bunching, the second-order correlation in Fig. 2c also shows suppressed emission for nearby adjacent pulses. This behavior is typically attributed to the existence of a metastable shelving state[36]. The suppression of emission in adjacent pulses disappears at higher pumping powers, as shown in Fig. 2d which is taken at an average pump power of 370 nW.

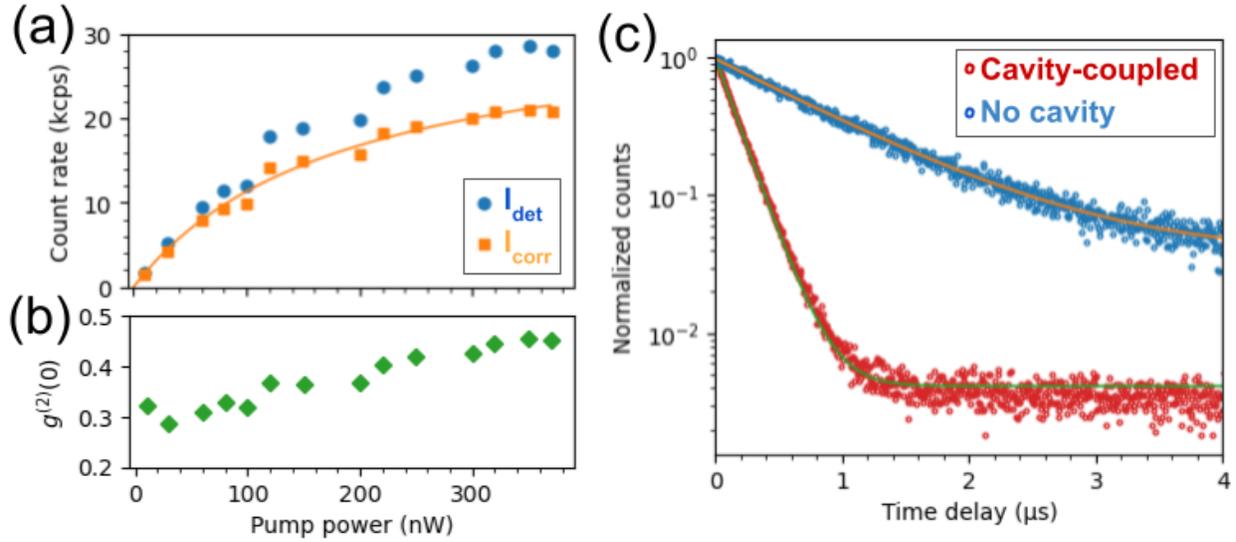

**Figure 3.** (a) Pump power dependence of detected count rates (blue circles) and $g^{(2)}(0)$-corrected count rates (orange squares). The orange line corresponds to a fit with a two-level atom saturation model. (b) $g^{(2)}(0)$ as a function of pump power. (c) Time-resolved lifetime measurements of the cavity-coupled T-center (red points) and a bare emitter outside the cavity (blue points). Solid lines represent fits to an exponentially decaying function plus a background.

We next quantified the brightness of the zero phonon line emission of the T center. Figure 3a plots the count rate measured as a function of the pump power. The emitter was excited using the same pulsed laser source that we employed for second-order correlation measurements. The blue dots represent the raw emission rates from the T center, which display a clear saturation behavior. These rates encompass both the emission from the T center and the background rates. To determine the count rate from the emitter exclusively, we applied $g^{(2)}(0)$ correction[37]. We measured the value of $g^{(2)}(0)$ for each pump power, as shown in Figure 3b. Using these data, we derived the $g^{(2)}(0)$-corrected count rates, represented by the orange dots in Figure 3a.

The solid line in Figure 3a is a numerical fit to the $g^{(2)}(0)$ corrected saturation curve based on a two-level atom model[38]. We fit the data to a function of the form $I(P) = I_{sat}P/(P + P_{sat})$, where $P$ is the pump power, and $P_{sat}$ and $I_{sat}$ are the saturation pump power and saturation emission intensity respectively. At the highest pump power of 370 nW, we obtained the $g^{(2)}(0)$-corrected count rates of 20.87 kcps. By comparing this count rate to the laser repetition rate of 200 kHz, we determined an overall efficiency of 10.4% from source to detector.

We performed a full photon budget of the collection and detection optics, as described in Supporting Information Section 1. From this photon budget we calculated the total efficiency into the lensed fiber to be 23.4%, and a total source efficiency of 63.4%. The reduced efficiency of the source is attributable to a number of factors including non-radiative decay, residual emission into the phonon sideband, and residual emission out of the left mirror. Additional inefficiency may arise from the fact that we are exciting the emitter incoherently, which means the emitter may emit in multiple polarizations. This problem can be rectified by employing direct resonant excitation of the emitter[39].

In Supporting Information section 2, we perform a similar analysis under continuous-wave excitation. From these measurements we determine a $g^{(2)}(0)$-corrected continuous-wave count rate of 236.3 kcps at the highest pump power. This count rate is more than two orders of magnitude larger than those previously reported using just a nanobeam without a cavity (1090 cps)[18]. These results further reinforce the significant emission enhancement achieved by the cavity.

The quintessential feature of coupling between a cavity and emitter is radiative lifetime enhancement, also known as the Purcell effect. To measure this effect we performed time-resolved lifetime measurements. Figure 3c shows the measurement results for the cavity coupled T center emitter (red circles), as well as the lifetime of a T center emission line located in a bare nanobeam without a cavity (blue dots). The solid lines represent a fit of the decay curve to a single exponential function plus background. From the fits, we determined that the cavity-coupled T center has an excited state lifetime of $\tau_c = 168.7 \pm 0.3$ ns, while the emitter in the bare nanobeam has a lifetime of $912 \pm 6$ ns. We performed similar lifetime measurements for an additional four T center emission lines in bare nanobeams (see Supporting Information Section 3). From the average of the five lifetime measurements we determine the average T center lifetime to be $\tau_0 = 838 \pm 7$ ns. From this value we calculated a lifetime enhancement of $R = \tau_c/\tau_0 = 5.0$.

To extract the Purcell factor, it is crucial to take into consideration the emission channeled into the phonon sideband. The Purcell factor is conventionally defined as the ratio of the radiative lifetime of the emitter within the cavity compared to that of a bare emitter. If an emitter displays only radiative decay, the Purcell factor directly corresponds to the observed enhancement in excited state lifetime. However, when dealing with emitters that also non-radiatively decay or decay into a broad phonon sideband, a substantially higher Purcell factor becomes necessary to attain the same level of lifetime enhancement observed in time-resolved lifetime experiments. To account for this effect, we derive a lower bound for the Purcell factor in Supporting Information Section 4 as a function of the lifetime enhancement and the Debye-Waller factor. From this lower bound, we find that the Purcell factor for the emitter-cavity system exceeds 18.

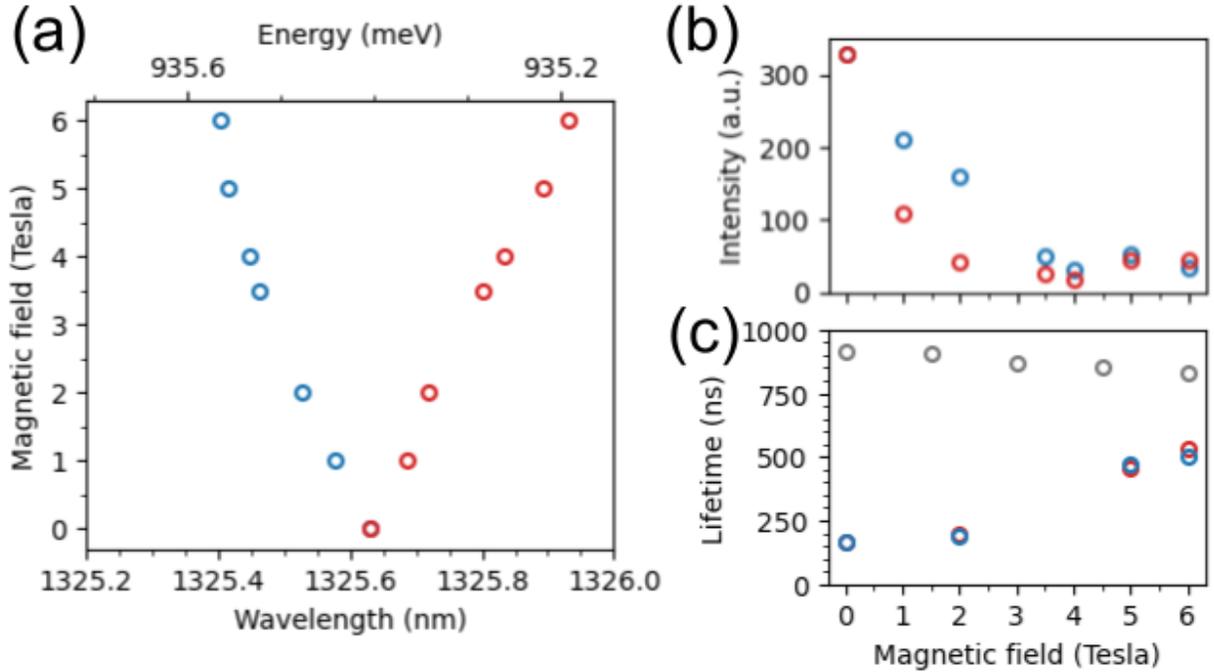

**Figure 4.** (a) Magnetic field dependence of the center wavelength of the Zeeman-split zero-phonon lines. Blue (red) circles correspond to shorter (longer) wavelength peaks. (b) Peak intensities as a function of the magnetic field. Blue (red) circles correspond to shorter (longer) wavelength peaks. (c) Measured lifetime as a function of magnetic field magnetic fields. Blue (red) circles correspond to shorter (longer) wavelength peaks, and gray circles are measured values for a T center in a nanobeam with no cavity.

To further demonstrate that the T center is coupled to the cavity, we applied an external magnetic field to the emitter. The field was applied in the plane of the nanobeam with the axis of the field oriented 45 degrees relative to the nanobeam direction. Applying a magnetic field has two effects. First, as we increase the magnetic field, we induce a Zeeman shift that splits the zero-phonon line resonance into a short and long wavelength branch. This wavelength shift tunes the emitter out of resonance with the cavity. Figure 4a plots the resonance wavelength of the two branches as a function of the magnetic field. We attained a 0.8 nm Zeeman splitting, sufficient to tune both branches out of resonance with the cavity. Second, the magnetic field changes the polarization of the dipole, which can further decouple it from the cavity. The change of the polarization strongly depends on the orientation of the T center relative to the nanobeam and applied magnetic field. These effects combine to modulate the coupling of the T center to the cavity mode.

Figure 4b-c plot both the intensity and measured lifetime as a function of the magnetic field for both the lower and upper branch. Both branches experience a significant reduction in brightness as we tune them out of resonance. Simultaneously, the lifetimes of the branches become longer and approach the lifetime of the bare emitters outside the cavity. The gray circles in Figure 4c show the measured lifetime of a T center emission line in a nanobeam without a cavity as a function of magnetic field. Unlike the cavity-coupled T center, the lifetime of the T center emission in the nanobeam does not change significantly as a function of the magnetic field. These results demonstrate that cavity coupling enhances both brightness and increases the spontaneous emission rate. Higher magnetic fields could continue to detune the emitter and extend the lifetime even further, uncovering the inherent lifetime dictated by non-radiative decay and coupling to leaky photonic modes. But our current tuning range is limited by the maximum magnitude of the magnetic field we can generate.

In conclusion, we demonstrated cavity-enhanced emission from the zero-phonon line of a single T-center coupled to a cavity. We attained a brightness enhancement exceeding two orders of magnitude and a fiber-coupling efficiency surpassing 23%. Furthermore, we demonstrated a fivefold decrease in the excited-state lifetime, corresponding to a Purcell factor greater than 18. Higher efficiency may be achievable through resonant-pulse excitation, ensuring deterministic preparation of the emitter in its excited state. Our measurements were conducted at 3.6 K, resulting in a phonon-broadened linewidth. By transitioning to lower temperatures and employing an isotopically purified substrate[15], we anticipate a significant reduction in the emitter linewidth, potentially enabling both high efficiency and indistinguishability. Ultimately, our findings constitute a significant step towards employing T centers as efficient, optically active spin qubits for application in quantum networking and distributed quantum computing.

AUTHOR INFORMATION

**Corresponding Author**

Edo Waks – Institute for Research in Electronics and Applied Physics and Joint Quantum Institute, University of Maryland, College Park, Maryland 20742, USA; Email: edowaks@umd.edu

**Author Contributions**

[#]F. I. and C.-M.L contributed equally to this work.

Acknowledgments

The authors would like to acknowledge financial support from the National Science Foundation(grants #OMA1936314, #OMA2120757, #PHYS1915375, and #ECCS1933546), the U.S. Department of Defense contract #H98230-19-D-003/008, and the Maryland-ARL QuantumPartnership (W911NF1920181).

# Supporting information for:

# Cavity enhanced emission from a silicon T center

*Fariba Islam,[†,‡,#] Chang-Min Lee,[†,‡,#] Samuel Harper,[†,‡,] Mohammad Habibur Rahaman,[†,‡,] Yuqi Zhao,[†,‡,] Neelesh Kumar Vij,[†,‡,] and Edo Waks[†,‡,\*]*

[†]Institute for Research in Electronics and Applied Physics and Joint Quantum Institute, University of Maryland, College Park, Maryland 20742, USA

[‡]Department of Electrical and Computer Engineering, University of Maryland, College Park, Maryland 20740, USA

## 1. Photon budget calculations

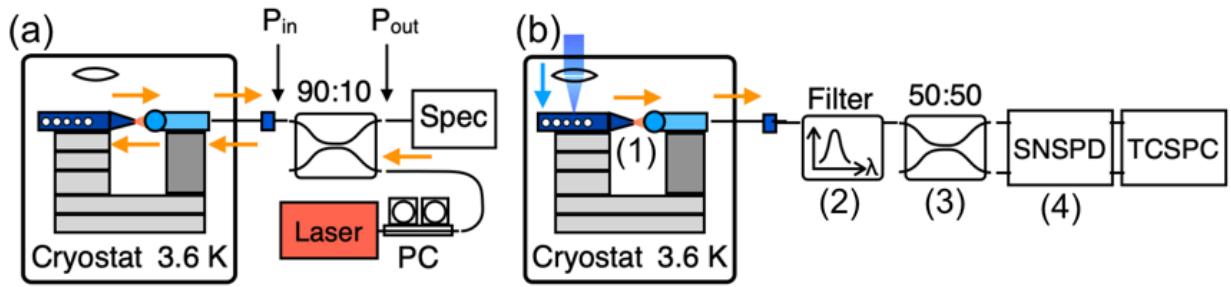

Figure S1: (a) Measurement setup for cavity reflectivity. PC: polarization controller, Spec: spectrometer. (b) Measurement setup for single photon emission. SNSPD: superconducting

nanowire single photon detectors, TCSPC: time-correlated single photon counter. Labels from (1) to (4) indicate optical components causing photon loss.

We measured the reflectivity of the nanobeam through the lensed fiber to characterize the cavity mode as well as a coupling efficiency $\eta_{fib}$ between the nanobeam and the lensed fiber (Figure S1(a)). We assumed that the coupling-in and coupling-out efficiencies are the same. The input and output power spectra were measured at the $P_{in}$ and $P_{out}$ positions, respectively. Assuming that the reflectivity of the photonic crystal mirror is unity, the off-resonance reflectivity is given by $R = P_{out}/P_{in} = \eta_{fib}^2 T_{fc}$, where $T_{fc}$ is the transmission of the 90:10 coupler (83%). From the reflectivity spectrum of Figure 2(a) in the main manuscript we obtained $R = 0,43$, and the corresponding fiber coupling efficiency $\eta_{fib} = 0.73$.

To assess the source efficiency, we performed photon budget analysis. Along the line of propagation of the single photon (Figure S1(b)), the transmission or detection/collection efficiencies at each components are: (0) cavity to nanobeam efficiency (from near-critical coupling) of $\eta_{nb} = 0.5$, (1) fiber coupling efficiency of $\eta_{fib} = 0.73$, (2) fiber spectral filter transmission of $\eta_{filter} = 0.58$, (3) 50:50 beam splitter and fiber cable transmission of $\eta_{bs} = 0.88$, and (4) SNSPD detection efficiency of $\eta_{det} = 0.88$. The $g^{(2)}(0)$-corrected detected count rates at the highest pump power was $I_{corr} = 20.87$ kcps with the repetition rate of the pumping laser of 200 kHz, which gives the overall efficiency of 10.4% (Figure 3(a) in the main manuscript). Using the optical components transmission values, we achieved efficiency to the lensed fiber $B_{fib} = (I_{corr}/R)/(\eta_{filter}\eta_{bs}\eta_{det}) = 23.4\%$, and the source efficiency of $B_{source} = (I_{corr}/R)/(\eta_{nb}\eta_{fib}\eta_{filter}\eta_{bs}\eta_{det}) = 63.4\%$.

## 2. Emitter Brightness Under Continuous Wave Excitation

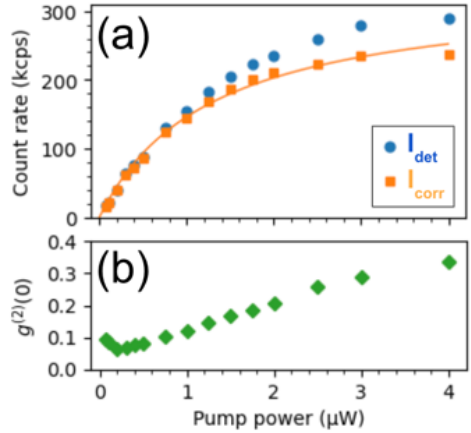

Figure S2: (a) Count rates as a function of pump power. Blue dots are raw data, and orange dots are $g^{(2)}$ corrected data. Solid-line represents a numerical fit. (b) Measured $g^{(2)}(0)$ as a function of pump power.

Figure S2 shows measured second-order correlation and brightness measurements. Figure S2(a) shows the photon count rates as a function of pump power. Blue dots represent raw data while orange dots represent $g^{(2)}$ corrected data. The $g^{(2)}(0)$ as a function of pump power is provided in panel b. This value remains below 0.5 for all measured data points. The solid line is a numerical fit of the $g^{(2)}$ corrected data to the same saturation curve used to generate Figure 3 of the main manuscript. At the highest pump power of 4 µW we attain a count rate of 236.3 kcps. These count rates are more than two orders of magnitude larger than the rates reported for a bare nanobeam, which achieved a maximum count rate of 1090 cps [1].

## 3. Lifetime Data for Emitters in a Nanobeam

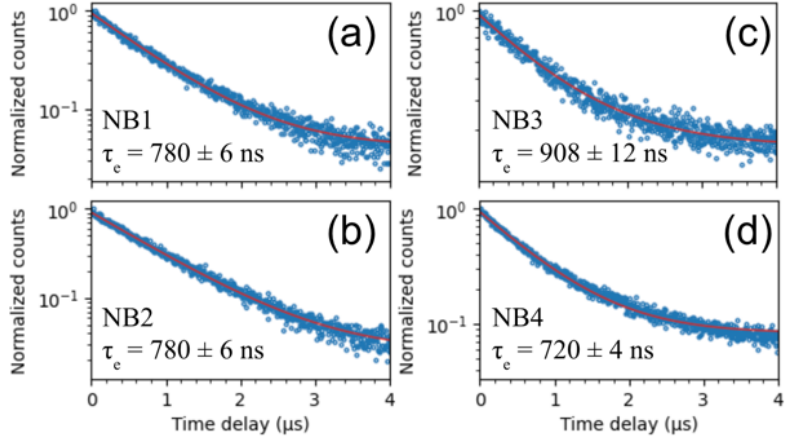

Figure S3: Lifetime measurements for four different T center emission lines in a nanobeam.

### 4. Calculation of the Purcell Factor

The excited state decay rate of a bare emitter, not coupled to a cavity, is given by

$$\gamma_0 = \gamma_z + \gamma_{nr} + \gamma_p$$

In the above equation $\gamma_z$ is the radiative decay rate into the zero-phonon line, $\gamma_{nr}$ is the non-radiative decay rate, and $\gamma_p$ is the decay rate into the phonon sideband. The decay of an emitter coupled to the cavity is given by

$$\gamma = \gamma_c + \gamma_{nr} + \gamma_p$$

where $\gamma_c$ is the decay rate of the emitter into the zero phonon line when coupled to the cavity. The lifetime enhancement is given by

$$R = \frac{\gamma}{\gamma_0} = \frac{\gamma_c + \gamma_{nr} + \gamma_p}{\gamma_z + \gamma_{nr} + \gamma_p}$$

We can rewrite the above equation as

$$R = \frac{F + \chi + \left(\frac{1}{D} - 1\right)}{\chi + \frac{1}{D}}$$

where $F = \frac{\gamma_c}{\gamma_z}$ is the Purcell factor, $\chi = \frac{\gamma_{nr}}{\gamma_z}$, and $D = \frac{\gamma_z}{\gamma_p + \gamma_z}$ is the Debye-Waller factor [2]. From the above equation we obtain

$$F = (R - 1)\left(\frac{1}{D} + \chi\right) + 1$$

When $R > 1$ we can therefore obtain a lower bound on the Purcell factor given by

$$F > \frac{(R - 1)}{D} + 1$$

Using the measured lifetime enhancement of $R = 5.0$ and the known Debye-Waller factor of the T center of $D = 0.23$ [3], we attain a lower bound on the Purcell factor $F > 18.4$.

*Quantum*, 1(2):020301.